\documentclass[]{spie}  

\usepackage{amsmath,amsfonts,amssymb}
\usepackage{graphicx}
\usepackage[colorlinks=true, allcolors=blue]{hyperref}

\title{The IRT Telescope on board the THESEUS mission}

\author[a]{Diego G\"otz}
\author[a]{Aline Meuris}
\author[a]{Eric Doumayrou}
\author[a]{Dehbia Lattab}
\author[a]{Frédéric Pinsard}
\author[a]{Samuel Ronayette}
\author[a]{Thierry Tourrette}
\author[a]{Herni Triou}

\author[b]{Bortolino Saggin}

\author[c]{Marco Giovanni Corti}

\author[d]{Stefano Covino}
\author[d]{Fabrizio Fronasiero}
\author[d]{Luca Oggioni}

\author[e]{Luca Terenzi}

\author[f]{Stéphane Basa}

\author[g]{Enrico Bozzo}
\author[g]{Ludovic Génolet}

\author[h]{Lauro Conti}
\author[h]{Mariachiara Celato}
\author[h]{Paul Hedderman}
\author[h,m]{Shaymaa Hussein}
\author[h]{Christoph Tenzer}

\author[i]{Karine Mercier}
\author[i]{Lander Ruiz de Ocenda}
\author[i]{Jean-Michel Le Duigou}
\author[i]{Adrien Fort}

\author[j]{Florent Robinet}

\author[k,l]{Andr\'as P\`eter Jo\'o}
\author[k]{Rebeka Kiss}
\author[k,l]{Bendeg\'uz Koncz}
\author[k,l]{L. Viktor T\'oth}

\affil[a]{Université Paris-Saclay, Université Paris Cité, CEA, CNRS, AIM, Orme des Merisiers Bât. 709, 91191, Gif-sur-Yvette, France}
\affil[b]{Università di Padova, CISAS “G. Colombo”, via Venezia 1, 35131 Padova (PD) Italy}
\affil[c]{Department of Mechanical Engineering
Politecnico di Milano Lecco, Italy}
\affil[d]{INAF / Brera Astronomical Observatory, Via Bianchi 46, Merate (LC), Italy}
\affil[e]{NAF - OAS Bologna, via P. Gobetti 101, I-40129 Bologna, Italy}
\affil[f]{Aix Marseille Univ., CNRS, CNES, LAM, Marseille, France}
\affil[g]{Department of Astronomy, University of Geneva, Switzerland}
\affil[h]{Institute for Astronomy and Astrophysics, University of T\"ubingen, Germany}
\affil[i]{CNES, 18 avenue Edouard Belin, Toulouse, France}
\affil[j]{Universit\'e Paris-Saclay, CNRS/IN2P3, IJCLab, 91405 Orsay, France}
\affil[k]{UoDebrecen: Faculty of Science and Technology, University of Debrecen, Bem tér 18, H-4026 Debrecen, Hungary}
\affil[l]{ELTE: Department of Astronomy, Eötvös Loránd University, Pázmány Péter sétány 1/A, H-1117 Budapest, Hungary}
\affil[m]{An-Najah National University, Nablus, Palestine}

\authorinfo{Further author information: diego.gotz@cea.fr}

\begin{document} 
\maketitle

\begin{abstract}
We present the Infra-Red Telescope (IRT), which is part of the payload of the THESEUS mission, one on the three phase A candidate missions for the M7 slot of ESA (launch date 2037). The IRT is a 0.7 m class telescope with an off-axis Korsch optical design, with imaging capabilities in the 0.7--1.8 microns range over a 15 x 15 arc min field of view. The IRT also provides slit-less low resolution spectroscopy (R$\sim$400) over a limited field of view of 2 x 2 arc min, in the 0.8--1.6 microns range. The goal of the IRT is to identify the near infrared counterparts to the Gamma-Ray Bursts (GRBs)  detected by the two other telescopes on board THESEUS (the XGIS and the SXI), and to measure on board its photometric redshift in near real-time. The position and the redshift will be transmitted immediately to ground to allow for deeper follow-up by the large telescopes (ELT, VLT, ...). If the source is bright enough, spectroscopy will be performed to characterize the GRB environment.
\end{abstract}


\section{Introduction}
\label{sec:intro}  

Gamma-Ray Bursts (GRBs) are fast transient sources initially detected in gamma-rays, where they last from a fraction of a second to several hundreds of seconds \cite{Zhang2019GRB}. 
Their origin has been mysterious for decades after their discovery in the late '60s of the last century \cite{Klebesadel1973GRB}, until the detection of a longer lasting panchromatic emission\cite{Costa1997Afterglow}, the so called \textit{afterglow}, following the gamma-ray \textit{prompt emission}. The afterglow is namely observable from X-ray to radio, from hours to weeks, or even months, after the GRB and carries peculiar information about the GRB distance, environment and nature.
It is thanks to the joint study of the prompt and afterglow emission, that we know today that gamma-ray bursts originate either from the merging of two compact objects (typically two neutron stars) or from the collapse of very massive stars ($>$50 times the mass of the Sun). The latter can be observed up to very large distances, when the Universe was very young. However, due to strong Lyman-alfa absorption of the light in the optical domain, in order to detect and study efficiently the GRB afterglows at redshifts (distances) beyond $z$=6 a telescope working in the near infra-red band is needed.

\begin{figure} [ht]
   \begin{center}
   \begin{tabular}{c} 
   \includegraphics[height=9cm]{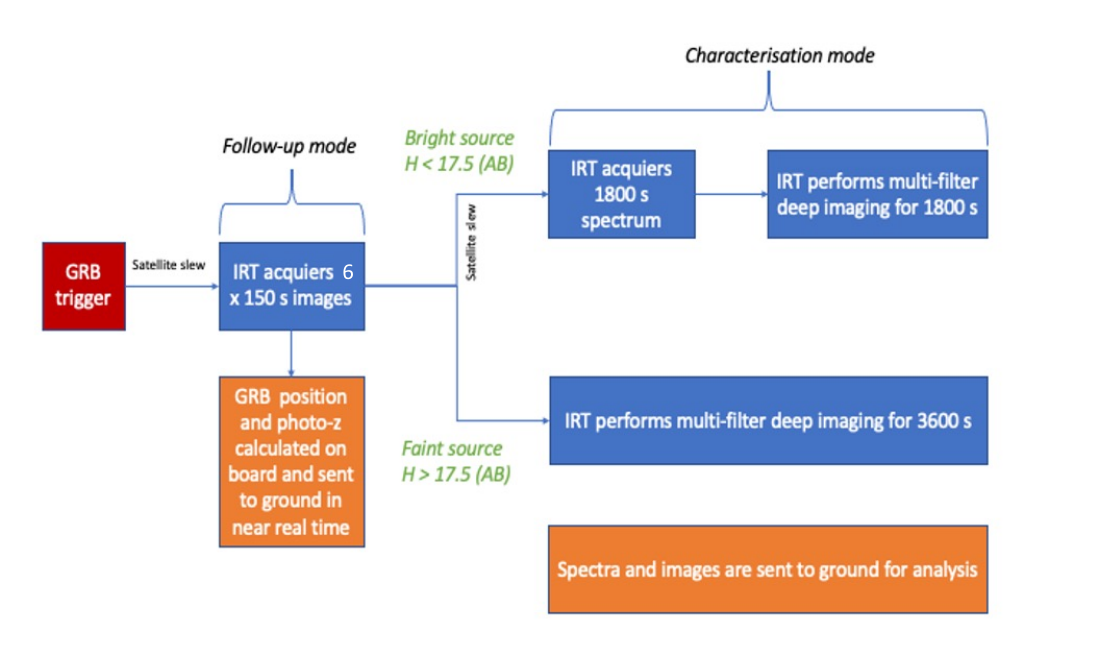}
   \end{tabular}
   \end{center}
   \caption[] 
   { \label{fig:operations} 
Overview of the IRT operational concept after a GRB trigger.}
   \end{figure}

\begin{figure} [ht!]
   \begin{center}
   \begin{tabular}{c} 
   \includegraphics[height=5cm]{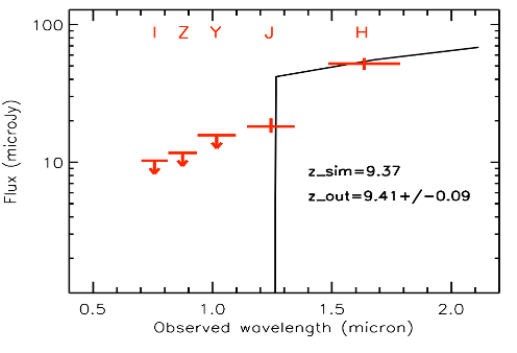}
     \includegraphics[height=5cm]{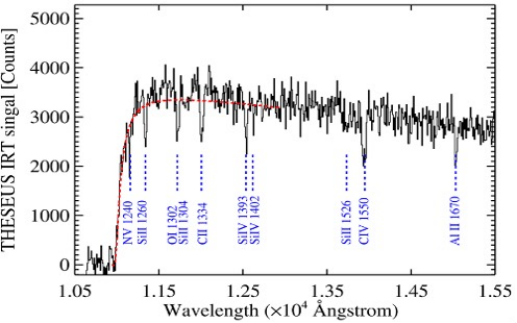}
   \end{tabular}
   \end{center}
   \caption[] 
   { \label{fig:exa} 
Left: Example of photometric fitting simulation for a GRB at z= 9.37. Right: Simulation of a GRB afterglow spectrum as observed by IRT.}
   \end{figure}

The IRT on board THESEUS \cite{amati} has been designed in order to be able to localize and measure the photometric redshift of THESEUS GRBs in near-real time. Following a detection by the X- and Gamma-ray Imaging Spectrometer (XGIS) or the Soft X-ray Imager (SXI) the satellite will perform a slew in order to put the GRB error region within the photometric field of view of the IRT. Then a first sequence is initialized, composed by a multi-filter observation lasting 6 $\times$ 150 s (co-adding individual frames of 25 s). At the end of this sequence, IRT is supposed to identify autonomously on board the GRB afterglow and measure its photometric resdhift, see Figure \ref{fig:exa} These results are transmitted to the ground immediately. Simulations show that in about 90\% of the cases IRT can reach an accuracy of better than 10\% on the redshift determination on-board, see Section \ref{sec:ics:analysis}.
Following this first sequence, if the afterglow is bright enough (H$<$17.5 AB), the satellite will perform a slew in order to put the source into the spectroscopic field of view (2 by 2 arcmin wide) for further characterization of the GRB and its environment, see Fig \ref{fig:exa}. In case the afterglow is not identified or it is fainter than H = 17.5 (AB), the imaging sequence will continue for 3600 s and the images will be sent to ground in order to be analysed. The IRT operational principle is schematised in Figure \ref{fig:operations}.

The main IRT characteristics are summarized in Table \ref{tab:chars}. There we report also the  limiting magnitudes for each filter, based on the requirements derived at mission level in order to detect the bulk of the GRB afterglow population.

   \begin{table}[ht]
\caption{Main IRT Charateristics.} 
\label{tab:chars}
\begin{center}       
\begin{tabular}{|l|l|}
\hline
Parameter & Expected Value  \\
\hline
Telescope Type & Focussing on-axis Korsch observing off-axis FOVs\\
\hline
Entrance pupil & 700 mm\\
\hline
Collecting area & $>$0.34 m$^{2}$\\
\hline
M1-M2 distance & 675 mm\\
\hline
Focal length & 6167 mm\\
\hline
Exit pupil & 35 mm\\
\hline
Pixel scale & 0.6 arcsec/pixel for an IRT detector pixel pitch of 18 $\mu$m\\
\hline
Fields of view & Photometry: $15^{\prime} \times 15^{\prime}$\\
& Spectroscopy: $2^{\prime} \times 2^{\prime}$\\
& Separation: 1.5$^{\prime}$\\
\hline
Wavelength range & 0.7--1.8 $\mu$m for photometry (filters I Z Y J H)\\
& 0.8--1.6 $\mu$m for spectroscopy\\
\hline
Photometric sensitivity (5$\sigma$, 150 s, AB) & I: 20.9 \\
& Z: 20.7\\
& Y: 20.4 \\
& J: 20.7\\
& H: 20.8 \\
\hline
\end{tabular}
\end{center}
\end{table}

\section{THE IRT System}
\label{sec:system}  

\subsection{Overview}

The IRT instrument can be decomposed into two independent systems. On one hand, the Instrument Optical System (IOS) is inserted in a cavity of the Payload module and surrounded by the Earth baffle provided by the satellite prime contractor; it converts infrared light into digital raw signals and requires a careful management of the optical and thermal environment to ensure the performance of the instrument. On the other end, the Instrument Control System (ICS) is a set of warm electronics boxes, possibly placed on a separate panel of the platform, to control all electronics functions and operations of the IOS and to perform the science data processing; it is designed to provide the necessary autonomy of the instrument for the on-board follow-up of gamma-ray bursts.

The IOS is divided in two main subsystems: 
\begin{itemize}
  \item The IRT Telescope, described in Section 3, collects and focuses light until the entrance pupil of the camera. The optical design is based on a Korsch off-axis telescope. The mechanical architecture is built around a reference optical bench: on one side, it holds the primary and secondary mirrors, on the opposite, it holds the folding and third mirrors and the field stops. 
  \item The IRT Camera described in Section 4 includes a Filter Wheel and Focal Plane to perform the wavelength band selection and the photon detection.
\end{itemize}

Electrical and mechanical design studies were done in phase A to define the architecture for the ICS, resulting in two separate boxes described in Section 5. The Electronics Control Unit (ECU) placed close to the Optical System deals with the pre-processing of the raw images and the control of the camera subsystems. It also implements several control lines for the critical items of the Optical System to achieve the mechanical alignments required for the image quality and the detector stability required for the optimal detector response. The Data Handling Unit (DHU) implements image processing and scientific on-board analysis to extract in close to real time the location of the source, its intensity and an estimate of the redshift. This compressed information can be rapidly broadcasted for joint observations with other ground based or space borne telescopes.

During the phase A, feasibility studies and trade-off analysis were led to address the following challenges: (i) Can we propose a mass and cost effective design for the telescope that guarantees a good stiffness at launch and the proper optical alignment in all orbital conditions of the THESEUS low-Earth orbit? (ii) Can we propose a thermo-mechanical design of the camera that satisfies in a small volume the various thermal constraints of its subsystems, with an acceptable cooling power to be provided by the cryocooler of the satellite integrator? (iii) Can we process on-board infrared images to detect and localise transient sources, even quite faint, and provide an estimate of the redshift to rapidly classify the sources and provide reliable alerts to the community?

   \begin{figure} [ht]
   \begin{center}
   \begin{tabular}{c} 
   \includegraphics[height=9cm]{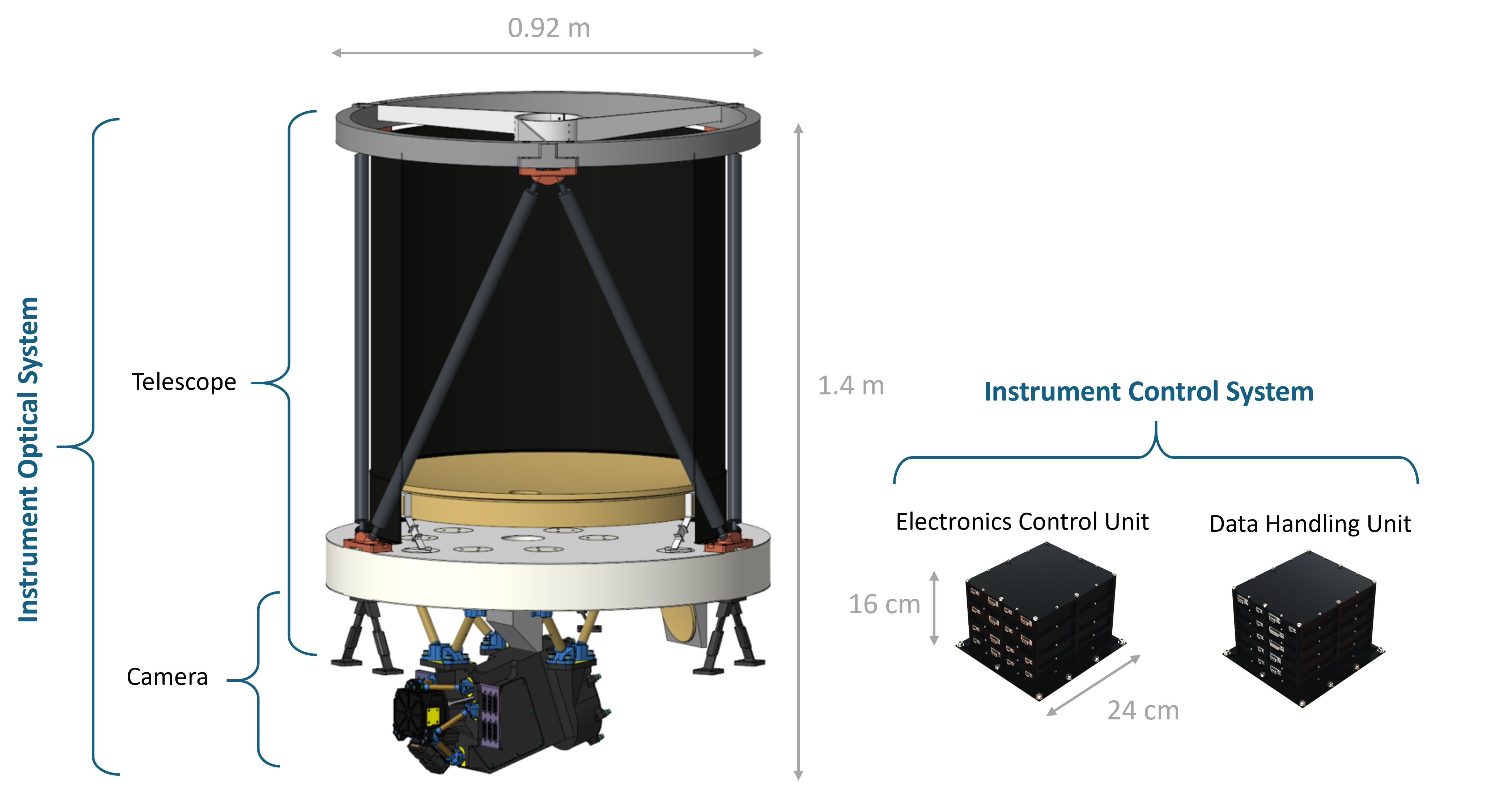}
   \end{tabular}
   \end{center}
   \caption[IRTsystem] 
   { \label{fig:IRTsystem} 
Overview of the main components of the IRT instrument decomposed in systems and subsystems. The dimensions without multilayer insulation are given to size the elements.}
   \end{figure} 
   
The results of these studies presented in the next sections led to a quite mature design at the end of phase A with the following system budgets (with classical 20 \% margins for this phase of the project):
\begin{itemize}
  \item A volume of a cylinder of 1 m diameter and 1.45 m height for the optical system mounted with an Earth baffle of 2.8 m. The ICS boxes represent a total volume of 16 L.
  \item A maximum power of 80 W, with main uncertainties coming from the thermal control budget.
  \item A mass of 123 kg for the telescope, 26 kg for the camera and 26 kg for the ICS, with main uncertainties coming from the mass of the harness (depending on the accommodation on the platform)
  \item A maximum telemetry budget of 2 Gbits during the follow-up phase (17 min) and 8 Gbits during the characterisation phase (1 hour).
\end{itemize}

\subsection{The IOS Thermal architecture}
The IRT IOS thermal design follows the general design split in two main components the telescope (TEL) and the camera (CAM).

\begin{figure} [ht]
   \begin{center}
       
   \begin{tabular}{c c} 
       \includegraphics[height=7cm]{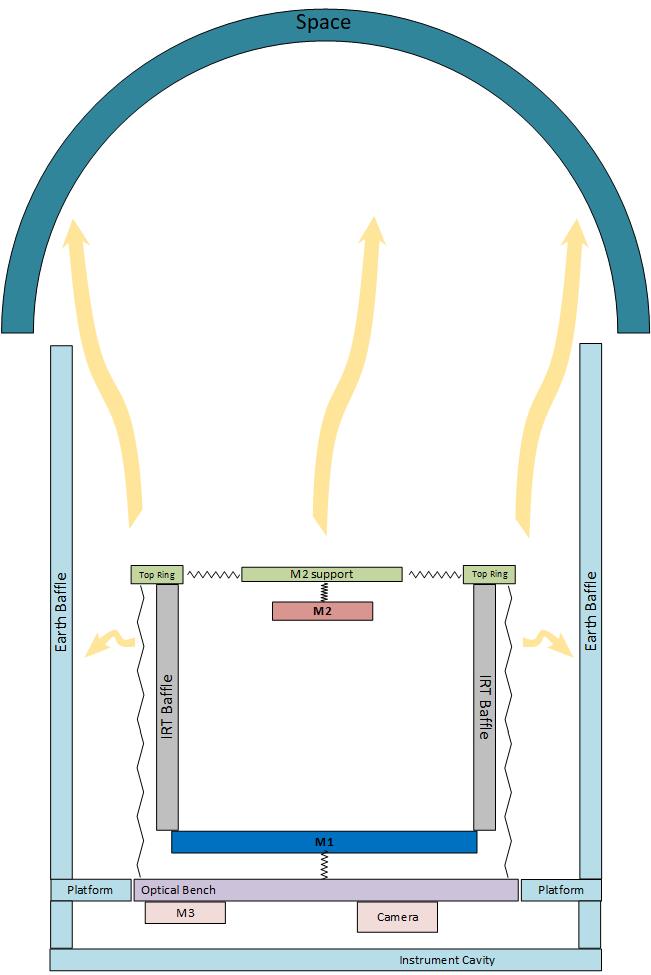}
       \includegraphics[width=0.5\linewidth]{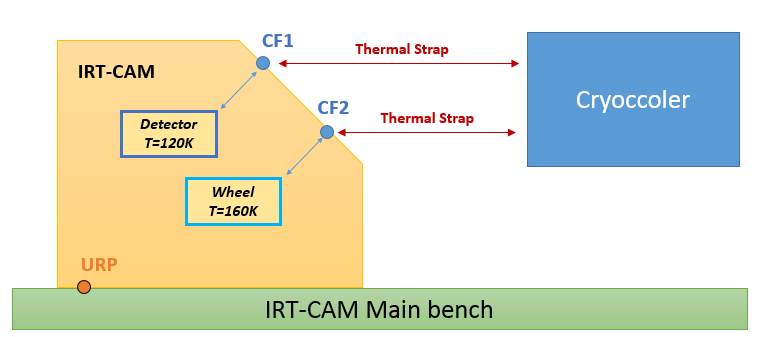}
   \end{tabular}
      \end{center}

   \caption[IRT-Thermal] 
   {\label{fig:IRThermal} 
Overall thermal concept of the IRT Optical system (left). The optical bench is the only conductive interface to the spacecraft, while the outer part is radiatively exposed to the space and the Earth baffle which is shielding the main optics from the other spacecraft surfaces and from Earth direct view. On the back of the optical bench the higher order optics and the IRT CAM are surrounded by the instrument cavity radiative environment. A focus on the camera thermal scheme and main interfaces is also shown (left).}
   \end{figure}

The thermal concept is summarized in the scheme at Figure \ref{fig:IRThermal}.

The main part of the telescope, external to the optical bench and instrument cavity, consists of the primary (M1) and secondary (M2) mirrors, which are surrounded by a cylindrical radiative shield (IRT baffle). The primary mirror structure is supported by the optical bench through bipods and shims. M2 support is connected through a 3-leg spider to a ring located on the top of the baffle. Such a top ring is supported by a truss structure to the optical bench. The thermal control scheme of the telescope is mainly passive, so no active conductive heat load is identified, with the exception of the thermal refocus actuators in the truss and a possible set of heating elements on the optical bench interfaces balancing excessive cooling during cold orbits. All the conductive links and radiative surfaces are designed with an insulating strategy with respect to external interfaces. Such an insulating concept allows a higher margin in the conductive interface to platform set 10 K higher with respect to the internal radiative environment in the instrument cavity
From the radiative point of view, the need of a thermally stable environment, preventing undesired optics temperature fluctuation on the orbit timescale, suggests the presence of a shield (Earth baffle) sized in order to avoid any thermal radiative flux from the Earth on part of the telescope assembly.

The thermal architecture of the camera relies on strict management of internal and external fluxes through three conductive interfaces known as Thermal Reference Points (TRPs), and one radiative interface. This strategy aims to protect sensitive components and efficiently dissipate heat from active elements.

 \begin{itemize}
\item TRP1 is the Detector Cold finger. The temperature is controlled by a Cryocooler and a control loop under ESA/prime responsibility to maintain the detector below 120 K to limit the dark noise. The requirement comes
from the state-of-the-art of detector performance.
\item TRP2 is the Structure Cold finger. The temperature is also controlled by a Cryocooler and a control loop
under ESA/prime responsibility to limit the thermal background of the camera. A first optical analysis indicated a temperature requirement of 160 K.
\item TRP3 is the main mechanical interface of the Camera, which is mounted on the Optical Bench of the
Telescope. The temperature of the Optical Bench can fluctuate between 200 and 240 K in all orbital
conditions, with 5 K stability along one orbit. The thermal design of the Camera shall be robust to these
thermal variations and maintain the alignment of the optical elements.
\item The radiative interface is also not controlled and ranges between 200 and 250 K.
The heat dissipating elements of the camera are the wheel during rotation ($\sim$ 50mW), the detector ($\sim$100mW), the FEE box ($\sim$ 500mW), the calibration unit when active ($\sim$ 500mW). The thermal design favours
the thermal conduction from these elements to the cold fingers, limits the conductive exchanges with the Optical
bench by means of the bipods in titanium and glass-fiber reinforced polymer (GFRP) and limits the radiative
exchanges with the environment by means of a multi-layer insulation surrounding the camera.
 \end{itemize}

\subsection{The IRT Electrical architecture}

The electrical architecture proposed for IRT shall implement all functional needs of the instrument, such as the thermal control of critical items identified in phase A (the refocusing system for the secondary mirror, the optical bench and the detector temperature), the control of the focal plane assembly, the control of the filter wheel assembly (including the motor and the wheel position sensor), the data acquisition and processing (covering housekeeping and science data), the interface communication with the on-board computer (including the instrument mode management). Note that the thermal control loop of the cryocooler is not part of the IRT electrical architecture.

\begin{figure} [ht]
   \begin{center}
   \begin{tabular}{c} 
   \includegraphics[height=10cm]{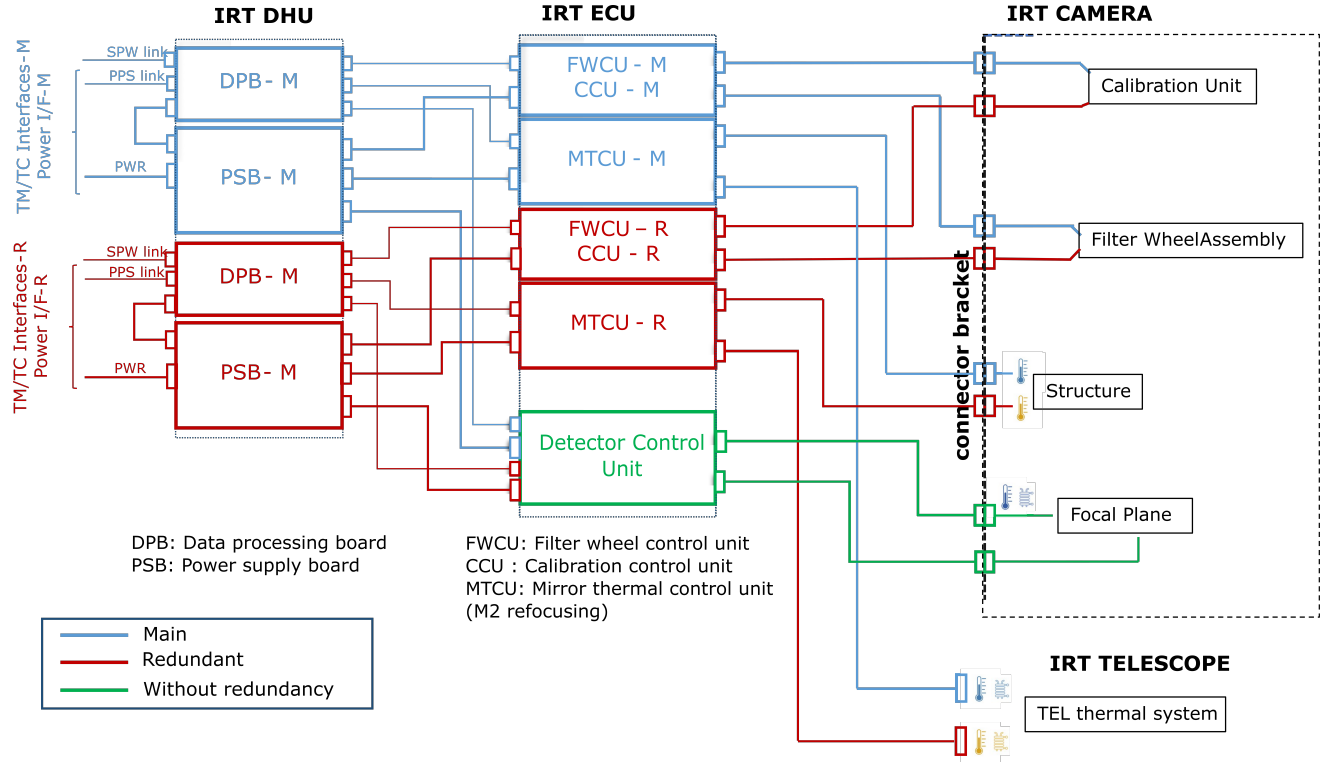}
   \end{tabular}
   \end{center}
   \caption[ElecArchi] 
   { \label{fig:Elec_Archi} 
Electrical architecture of the IRT instrument.}
   \end{figure}

The resulting architecture is presented in Figure \ref{fig:Elec_Archi}. The Data Handling Unit (DHU) is in interface with the platform includes a Data Processing Board (DPB) for the scientific data processing and for the management of the modes, tele-commands and data packets transfer, and a Power Supply Board (PSB) to generate secondary voltages to the control units. The Electronics Control Unit (ECU) is a second electronics box, closer to the Optical System, for the control of the telescope (M2 mirror refocusing system and possibly the thermal control of the optical bench) and for the control of the camera (filter wheel, focal plane, calibration unit). All boards of the IRT Instrument Control System (DHU and ECU) are duplicated for cold redundancy except from the Detector Control Unit driving the Focal Plane Assembly (redundancy would add more risk and complexity than benefits given that the FEE box has no redundant connectors).

\section{The IRT Telescope}

\begin{figure} [ht]
   \begin{center}
   \includegraphics[height=6.7cm]{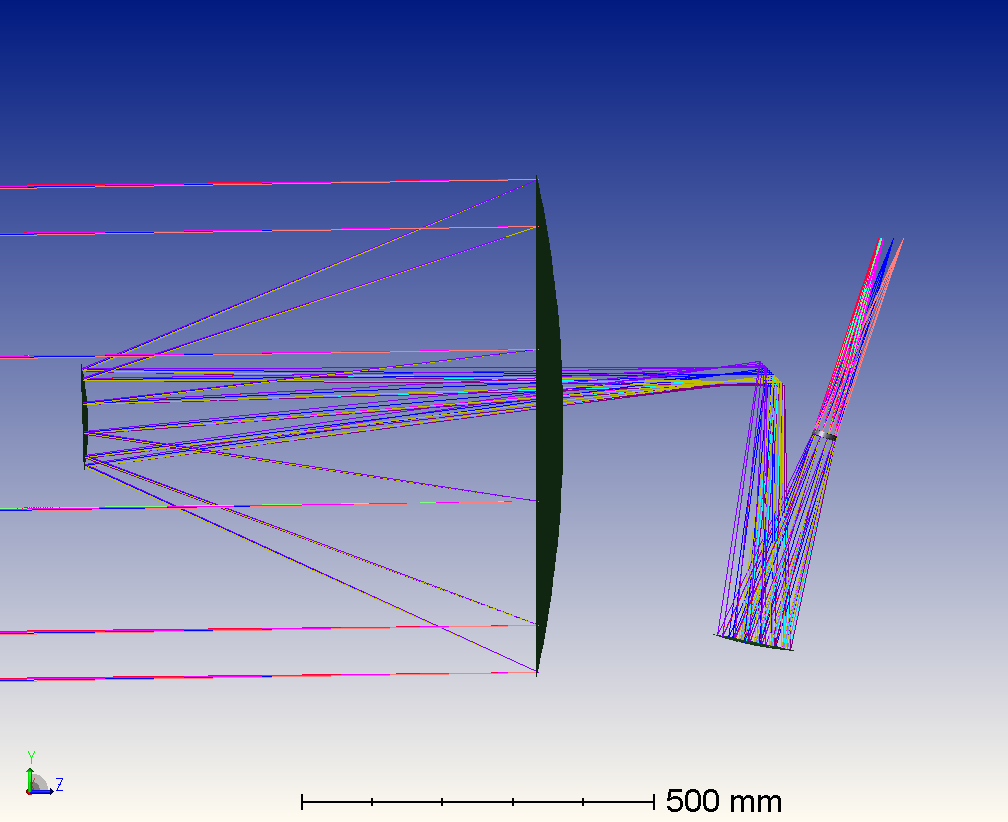}
      \includegraphics[height=6.7cm]{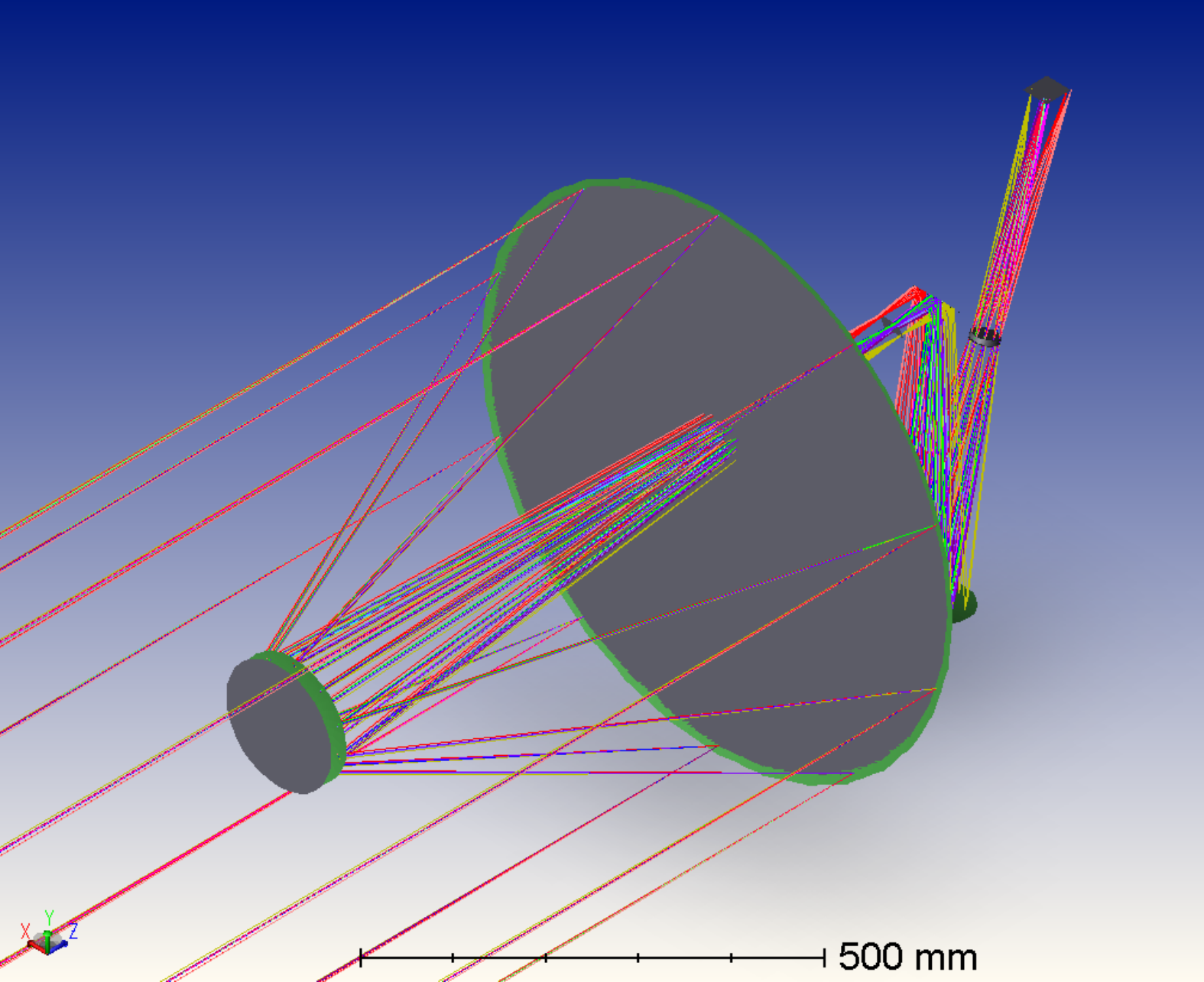}

   \end{center}
   \caption[] 
   { \label{fig:opt} 
IRT Telescope optical scheme.}
   \end{figure}

The IRT telescope is based on a Korsch optical scheme, with a primary mirror diameter of 700 mm and a focal length of about 6m. Its optical scheme is depicted in Figure \ref{fig:opt}.

The telescope structure has been implemented according to the scheme reported in Figure \ref{fig:IRTsystem}. The telescope mirrors are made of Zerodur to achieve dimensional stability over a the large temperature range, expected in thr different operational conditions. The main telescope structural components determine the relative positioning of the mirrors are the M2 truss and the optical bench. The six beams of the M2 truss are made with a CFRP tube with glued INVAR end fittings and have been designed to achieve an ideally zero global CTE along the beam by exploiting the negative CTE of the CFRP portion. The zero CTE was developed considering a totally passive system, nevertheless, the current baseline foresees to include, on each beam, a titanium insert whose temperature is actively controlled with a film resistance electrical heater. The truss in this configuration is acting as a thermal actuator allowing us to actively change the M1-M2 distance, a function identified as the refocusing system. The truss structure supporting M2 was preferred to a structural baffle because it can be radiatively insulated while the baffle, at least internally, must be an optical absorber and is therefore directly influenced by the external radiative fluxes, mainly generated by the Earth albedo. The optical bench is a sandwich structure made by two CFRP skins and an aluminium honeycomb. The latter is the only element in the optical system with a large CTE, making the M2-folding mirror distance temperature dependent. Nevertheless, it has been verified that the effect on the optical performances can be compensated with the M1-M2 distance tuning, the residual effect being a tilt on the telescope line of sight that can be managed with the data processing. From a thermal point of view the telescope system has been insulated as much as possible from the environment, i.e. the THESEUS satellite and the Earth that are the two sources of heat fluxes. The reduction of the Earth fluxes is the most critical issue because the telescope large aperture is behaving like a black body due to the baffle that should behave as a perfect absorber. The only way to reduce the fluxes is using an external shielding, this is achieved with an almost 2 m long external baffle, the Earth baffle. 
In the following a more detailed description of the main telescope structural elements.


{\bf The Earth baffle} is part of the satellite, structurally and thermally de-coupled with the telescope, but from the thermal point of view it has a strong impact because of the unavoidable radiative coupling, so, the telescope thermal model must include it. The main components of the mechanical system are identified in Figure \ref{fig:IRTsystem}, following the load path they are analysed in the following.

{\bf The mounting bipods} are the interface between IRT and the S/C. They provide thermal insulation from the S/C, implement an isostatic constraint scheme but achieving the stiffness and strength required to match the requirements about the lower eigenfrequency and the applied loads. The current design of the bipods is based on a titanium structure. The thermal conductance figure is not a driver for the average telescope temperature because as baseline the S/C has a temperature range not far from that of the telescope but, the insulation is relevant to smooth the temperature transients generated in the worst case of repointing from an hot to a cold condition that might be requested to track a newly identified source. 

{\bf The Optical bench, (OB)} is the main structural element of the telescope to which all the optical elements are connected. Besides holding the telescope optical chain, the OB is the interface with IRT camera. The structural requirements for the OB are transferring and withstanding the loads coming from the interface bipods to the mirrors and IRT-CAM mounting supports and providing stiffness to match the requirement on the first natural frequency. The stability of the telescope optical system has led to a sandwich structure with low CTE CFRP skins and an aluminium honeycomb. Aluminium inserts are foreseen to interface the OB with S/C and IRT-CAM bipods, M2 truss end fittings, M1, M3 and FM holding structures and the internal baffle. 

{\bf The M1 mounting structure} is a bipod system conceptually equivalent to the S/C interface bipods and with analog requirements. The peculiarity is that these bipods interface directly with the Zerodur material of M1, so, to match at best the CTE, the bipod structures are made in INVAR. The most critical requirement for this structure is the compliance in radial (with reference to the M1 disk) direction that determines the thermo-elastic loads at the interface. The conflicting requirements are stiffness/strength.

{\bf The M2 truss structure} must warrant a stable position of the M2 mirror wrt M1. The dimensional stability requirement has led to a CFRP truss with INVAR end fittings. Once again, the kinematic scheme is “the three bipods” one, providing an isostatic configuration for the mounting of the M2 spider. The almost 700 mm M1-M2 distance leads to long bars whose axial stiffness is driving the first eigenfrequency. Currently it has been designed with the requirement to stay above 60 Hz, but it can be tuned to avoid coupling with the S/C modes once these will be available.

{\bf The M2 spider structure} is within the telescope FOV therefore, beside the usual structural performances, its main requirement is minimizing the obscuration i.e. the blades thickness. The current design is based on a CFRP laminate. The design driver is the stiffness to maintain the first natural frequency above that of the truss. The internal surfaces are black coated (Aeroglaze Z306 TBC) to minimize the reflections on the blades. The M2 mirror is a lightened Zerodur structure that glued in the central part to an invar interface providing radial flexures. The M2 assembly will be glued to the spider once the alignment procedure will be completed.

{\bf The Internal baffle} is a plain tube with black coating (Aeroglaze Z306 TBC) on the internal side. With the current instrument envelope there would be no space to add internal vanes at least in the mid height region where the truss is closer to its surface. The baffle is a CFRP laminate skin with the only structural requirement of supporting itself. It is mounted on the optical bench with compliant and thermal insulating mountings. The outer surface is a low emissivity vacuum deposited gold (VDG) to minimize the disturbance on the truss.

\begin{figure} [ht]
   \begin{center}
   \begin{tabular}{c} 
   \includegraphics[height=5cm]{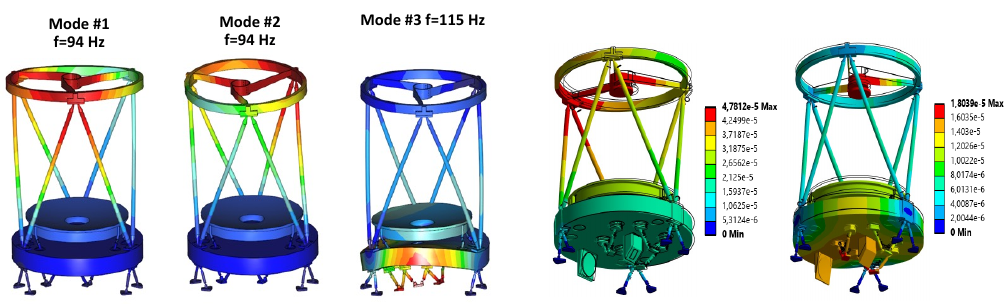}
   \end{tabular}
   \end{center}
   \caption[] 
   { \label{fig:deformations} 
Left: the three lower modes from modal analysis. Right deformations due to the gravity release.}
   \end{figure}

The whole structure has been sized through finite element method (FEM) modelling in order to fulfil the stiffness and strength requirements, the first depending on a first natural frequency above 60 Hz and the latter with a design load of 40 g and considering the expected random vibration spectra expected at the mounting interface. Beside the sizing of the structures, FEM models have been used to determine the optical system deformations generated by different mission phase such as the gravity release after launch of the temperature distributions generated in different operational scenarios. In Figure \ref{fig:deformations} we show some examples of the analyses results. 

   \subsection{Telescope Thermal Control}

\begin{figure} [ht]
   \begin{center}
   \begin{tabular}{c} 
   \includegraphics[height=6.8cm]{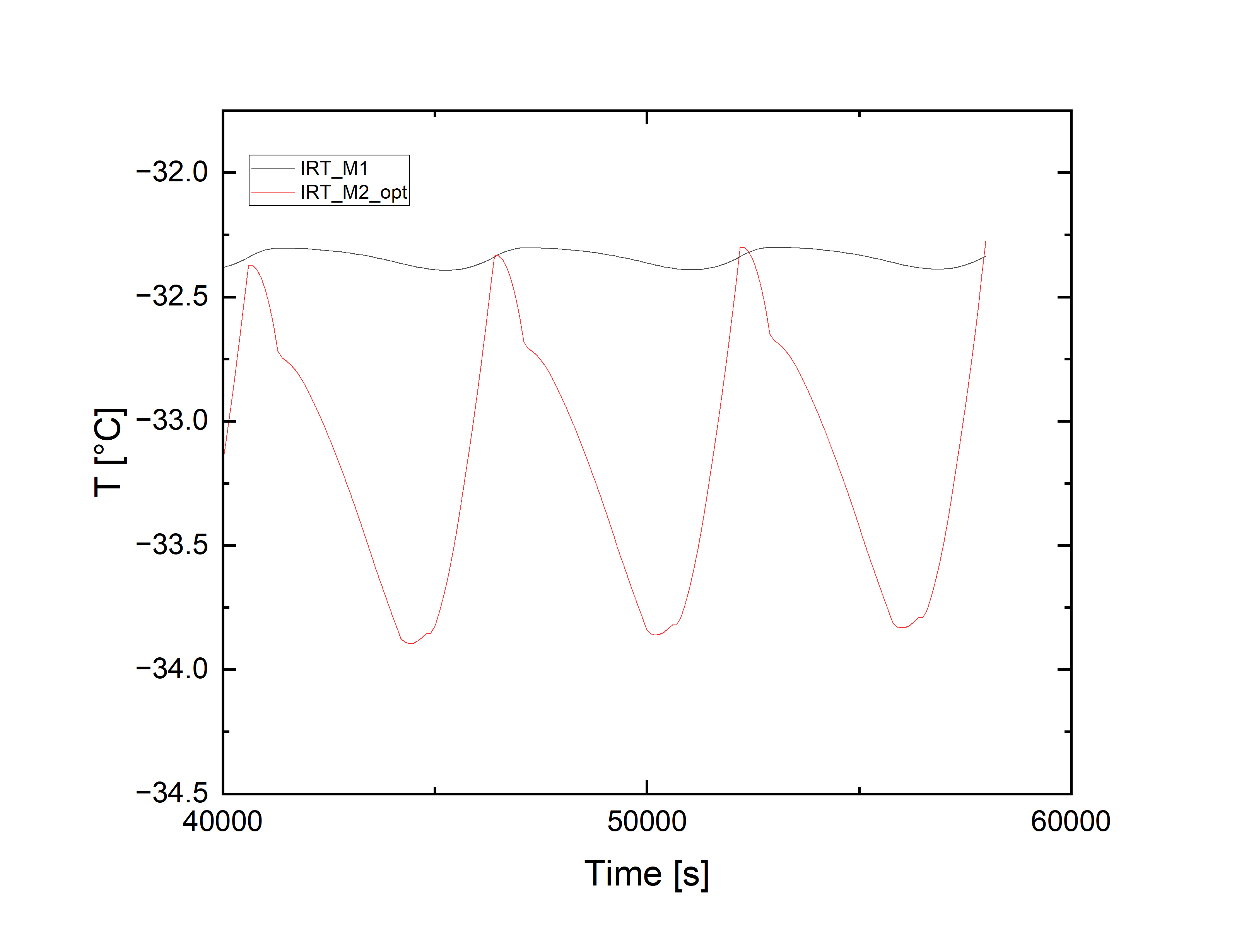}
\includegraphics[height=6.8cm]{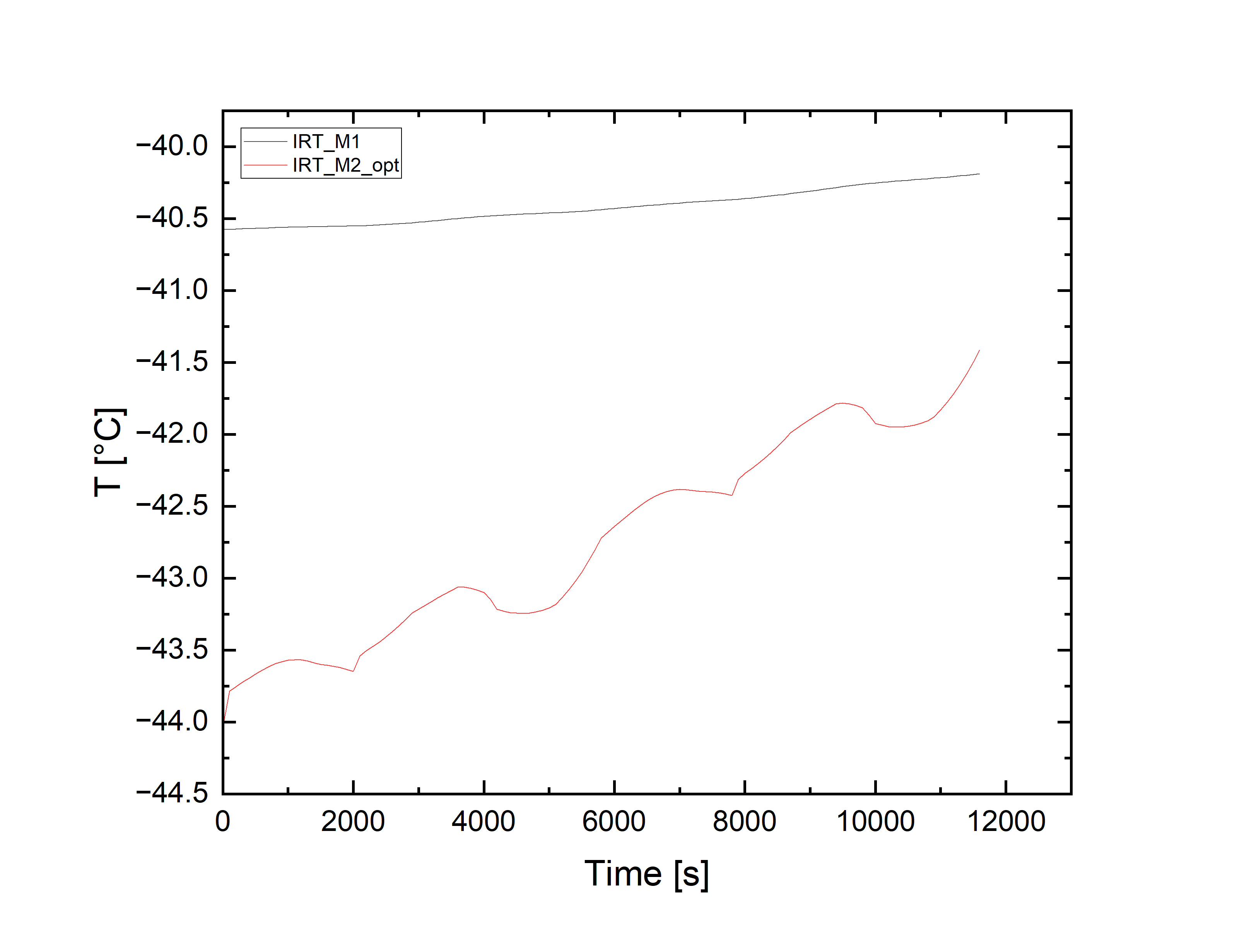}
   \end{tabular}
   \end{center}
   \caption[] 
   { \label{fig:thermal} 
Temperature predicted in orbital conditions, left the hottest condition, right the transient after a pointing change.}
   \end{figure}

   The IRT telescope optical quality depends on the thermal distribution on the optical elements and structural components connecting them. The ideal condition would be a uniform temperature at about 240 K to reduce the self-emission to a negligible contribution in the stray-light budget. The 240 K is also the requirement of the IRT-CAM conductive interface. For thermal emission temperatures lower than 240 K would be welcome, nevertheless wide operative temperature ranges become challenging for the Thermo-Elastic distortions, so currently the average lower temperature is limited to about 220 K. The actual temperature distribution on the telescope depends on the environmental fluxes interface temperatures and thermal control design. The telescope thermal control concept is based on a system as much isolated as possible from the thermal disturbances coming from the environment and the interface with the S/C. The main disturbance input is the telescope aperture that is facing mostly the Earth baffle and, depending on the orbital conditions, the Earth or deep space. This aperture cannot be managed in terms of emissivity/absorptivity. To reduce stray-light in fact, all internal surfaces, except for the mirrors, are black painted. The main radiative link is with the telescope internal baffle that is conductively and radiatively insulated from the telescope structure. The internal baffle exhibits large temperature changes, depending on the orbital fluxes so, its external surface has a low emissivity coating to reduce the radiative disturbance on telescope structures. The telescope external surfaces are covered with MLI. Conductively, the mounting bipods are titanium made, to provide for thermal insulation. The developed design can be completely passive nevertheless analysis have been carried out also to evaluate the power needed for thermal actuators of the refocusing system and to control the optical bench temperature. The power needed in the worst case has been estimated for each control in the range of 3 W, so with an acceptable impact on the telescope power budget. In the Figure \ref{fig:thermal} some examples of the temperature evolution of the key optical components M1 and M2 during the hottest orbital phase and during a transient from two orbital conditions.

\subsection{Telescope Manufacturability}

The requirements regarding the optical performance have been verified considering also the achievable quality of the optical system. For the optical components the tolerances on both curvatures and higher order shape errors have been considered. For the positioning accuracy state-of-the-art assembling procedures, based on the geometrical characterization of all components, have been considered. Finally, three degrees of freedom have been assumed as compensation in the alignment procedure and accuracy for each position has been assumed considering the achievable resolution of micro-positioning systems. The combined effect of all errors has been analysed using a Monte Carlo process and the results are summarized in the plot of Figure \ref{fig:eed}. 

\subsection{STOP analysis}

\begin{figure} [ht]
   \begin{center}
   \begin{tabular}{c} 
   \includegraphics[height=8cm]{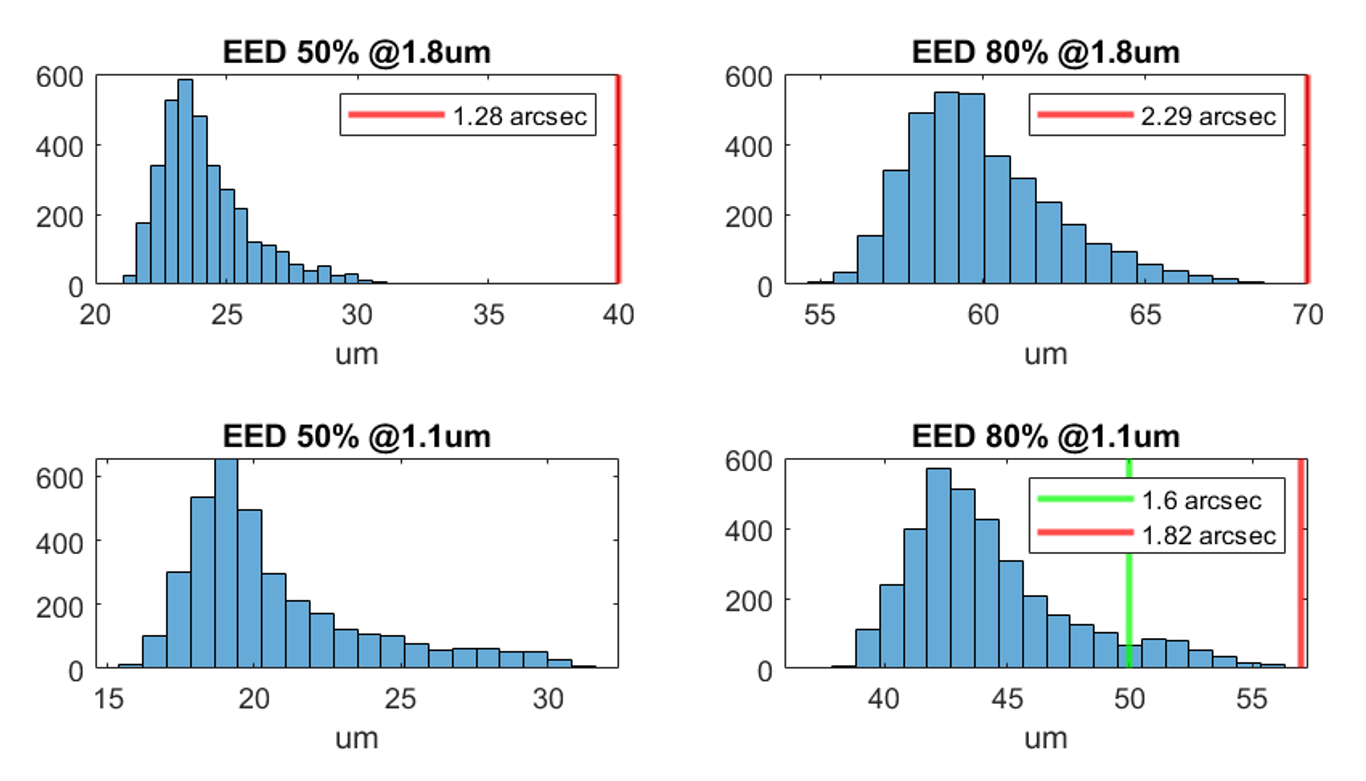}\\
\includegraphics[height=8cm]{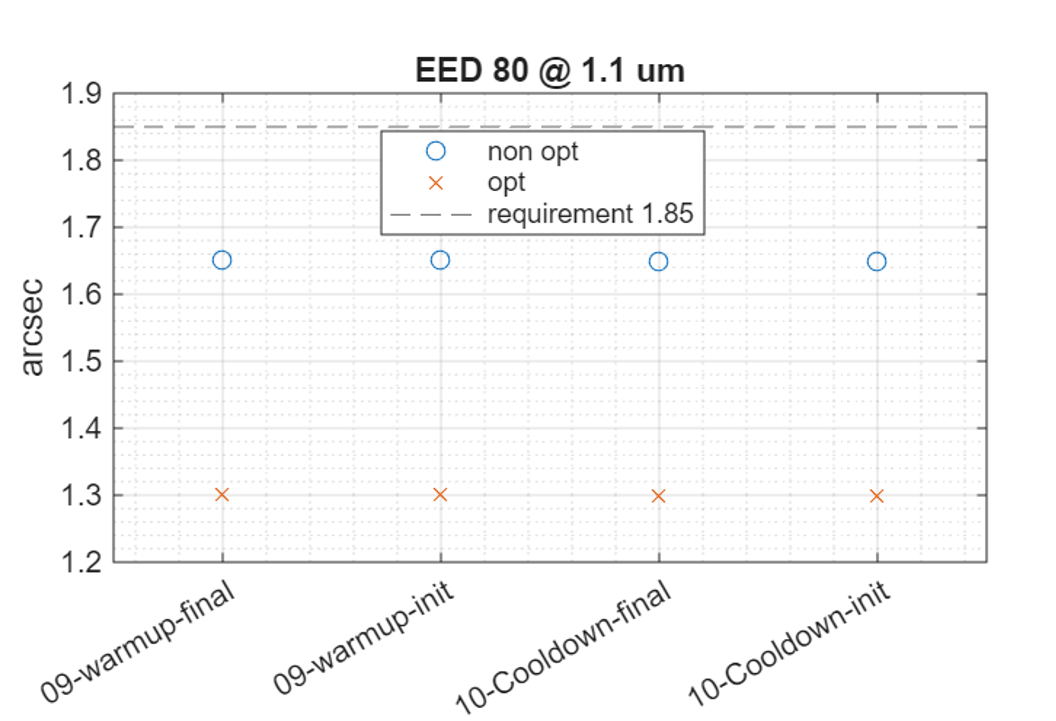}
   \end{tabular}
   \end{center}
   \caption[] 
   { \label{fig:eed} 
Top: frequency distribution of the Monte Carlo analysis of the optical performances including manufacturing and alignment tolerances. Bottom: the optical performances during the repointing transients.}
   \end{figure}

The structural thermo-optical analysis (STOP) has shown that optical performances might be achieved in all environmental sizing conditions even without changing the M1-M2 distance. The refocusing system has been maintained in the baseline configuration to increase the robustness of the system with respect to difficult-to-predict post launch misalignments. The refocusing system is thermally actuated, therefore thermal analyses have been carried out also including the thermal actuator. STOP analysis considering an extension of the S/C conductive interface temperature range has evidenced that in the extreme cold case with S/C interface at 200 K the optical quality would be achieved but with little margin. In case this interface condition was confirmed a thermal control channel would be used to limit the optical bench lower temperature above 225K i.e. the minimum temperature of the nominal cold case. In this way the telescope optical quality margins can be fully restored even with the extended range of the interface temperature. In Figure \ref{fig:eed}, on the right, an example of the analysis showing the effect of the worst transients during a satellite re-pointing showing that in all cases the requirements are met. 

\section{The IRT Camera}
\begin{figure} [ht!]
   \begin{center}
   \begin{tabular}{c} 
   \includegraphics[height=7cm]{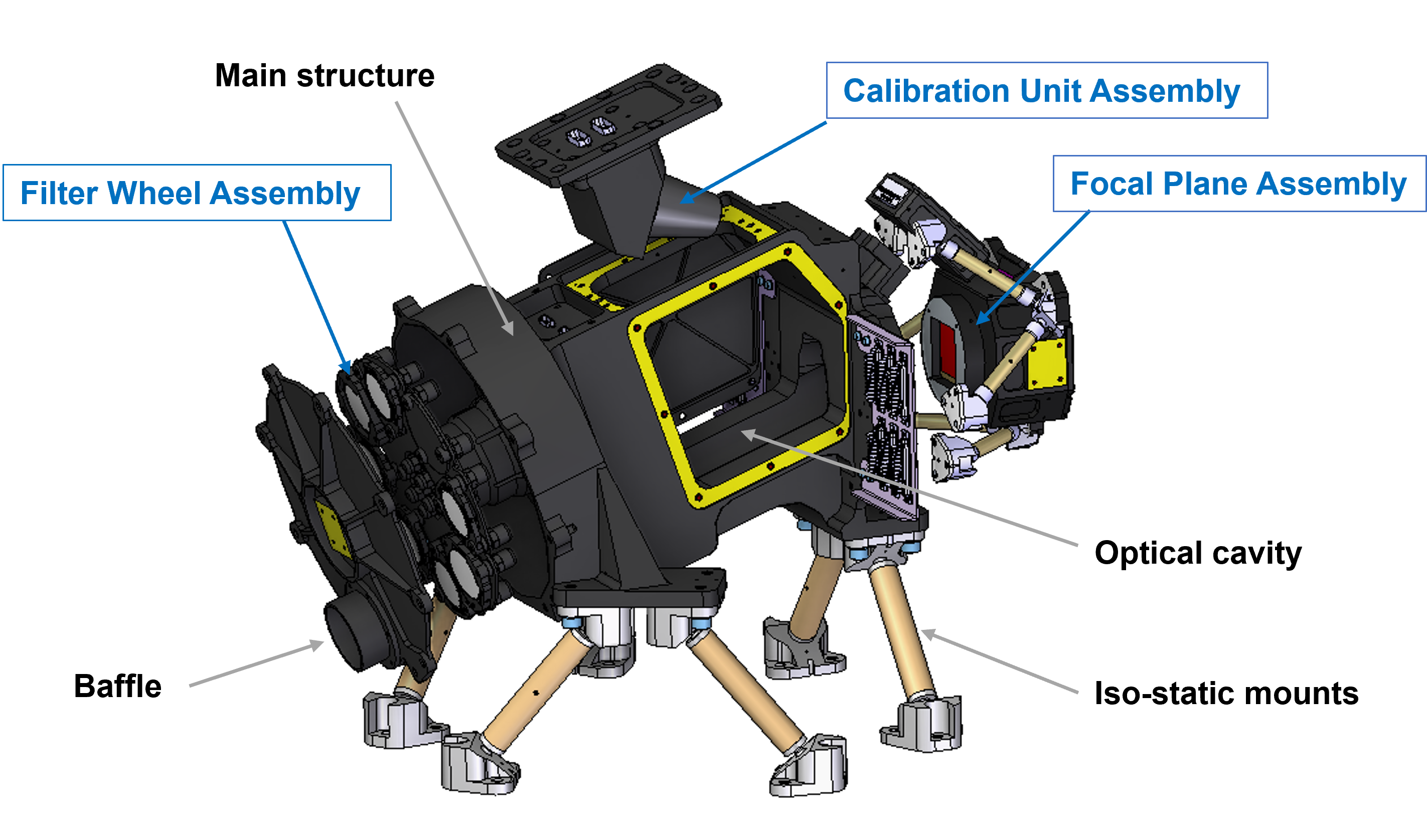}
   \end{tabular}
   \end{center}
   \caption[IRT-CAM] 
   { \label{fig:IRT-CAM} 
Exploded view of the IRT camera showing the main functional sub-assemblies and the mechanical and structural elements to combine them.}
   \end{figure}

The IRT Camera visible in Figure \ref{fig:IRT-CAM} implements in a compact volume three functional sub-assemblies : a Filter Wheel Assembly (FWA), a Focal Plane Assembly (FPA) and a Calibration Unit Assembly (CUA). These elements are mounted in a structural assembly made of aluminium alloy (Al6061). This material choice is the result of a phase A trade-off analysis: in comparison to silicon carbide known for its excellent stiffness and thermal stability, it allows more flexibility in the design, a closed structure for a good stray-light management and the integration of the Calibration Unit inside the optical cavity. The Camera is mounted on the Optical Bench of the telescope by means of iso-static mounts (3 bipods) to guarantee a good thermal insulation and immunity to the telescope temperature variations, if any. A detailed justification of this structure material and of the full design of the camera is available in this Volume \cite{camerapaper}. This section focuses on the main features of the design linked to the top-level requirements.

\subsection{The Filter Wheel Assembly} 

The Filter Wheel includes 8 positions : 5 bandpass filters (I-filter for the 0.73-0.88 $\mu m$ band, Z-filter for 0.82-0.98 $\mu m$ band, Y-filter for the 0.96-1.08 band $\mu m$, J-filter for 1.12-1.33 $\mu m$ and H-filter for the 1.48-1.78 $\mu m$ band), one grism (0.8-1.6 $\mu m$), one thermal filter for wide-band observation (0.8-1.8 $\mu m$) and one closed position for the Calibration, Off and Safe modes of the camera. The mechanical assembly is a heritage of the EUCLID NISP instrument. It foresees a wheel made of Invar with a direct drive stepper motor for a simple and reliable mechanism. The selected motor is the SAGEM 23pp63 04 03 01 reference, compatible with cryogenic temperatures and similar to references qualified for Euclid for the VIS instrument shutter and the NISP instrument. Optical mounts designs including gluing processes of the silica filters are based on the NISP design. Specific developments for the THESEUS mission were done in phase A for a multiple-position wheel position sensor, including prototyping.

\subsection{The Focal Plane Assembly}

The Focal Plane Assembly is based on the 2048 $\times$ 2048 pixel HgCdTe Hawaii-RG detector from Teledyne. The cut-off wavelength is planned to be optimised to 2.3 $\mu m$ as a good compromise between dark current and thermal background rejection. The 18 $\mu m$ pixels correspond to a pixel scale of 0.6 arcsec with the IRT optical design, leading to a maximal field of view of 20 $\times$ 20 arcmin$^2$. The active area of the detector is actually divided into two areas thanks to two field stops placed on the optical path: (i) a 15 $\times$ 15 arcmin$^2$ field of view for photometry used during the Follow-up, the Deep Imaging and the Guest Observer mode (ii) an area for the spectral dispersion of the 2 $\times$ 2 arcmin$^2$ field of view used during the Characterisation mode of bright sources (in Grism position). The H2RG is controlled and read out by the SIDECAR front-end ASIC being the most mature technology for this specific device. The packaging qualification was performed for the JWST and the Euclid mission down to 130 K. The detector temperature requirement is a maximal temperature of 120 K to limit the dark current to 1 or 2 electrons/pixel/second. In order to limit heat loads requirements on the cryocooler that will bring this power, the design of the camera foresees to move the Front-end Electronics box outside the detector housing with a thermal decoupling (and a thermal coupling to the Camera mechanical structure). 

\subsection{The Calibration Unit Assembly}

The Calibration Unit Assembly is embedded in the IRT Camera to perform flat field illumination, measure pixel response linearity, verify quantum efficiency uniformity during the mission after detector ageing by displacement damage dose. If the optical concept is a heritage of the NISP instrument (LED with a reflector and baffles), the mechanical design is fully specific of IRT due to the very small volume. Moreover, new calibrated infrared sources were investigated to be independent of the Euclid Russian procurement source. The phase A study results in a mechanical and optical design fully compliant with the technical requirements and a list of LED from three to four alternative procurement sources covering the wavelength range of IRT to be characterized and qualified for space applications. 

\section{The IRT Instrument Control System}
The IRT Instrument Control System (ICS) comprises the warm electronics which powers, controls, and processes telemetry from the various electrical subsystems of the IRT IOS. Originally envisioned as a single unit, the ICS underwent a significant redesign study in phase A, in order to optimise the thermo-mechanical performance of the system. This study resulted in the ICS being split into two distinct electronics units: the Data Handling Unit (DHU) and the Electronics Control Unit (ECU). Both ICS units are mounted internally in the service module (SVM) of the THESEUS platform.

\begin{figure} [ht]
   \begin{center}
   \includegraphics[height=7cm]{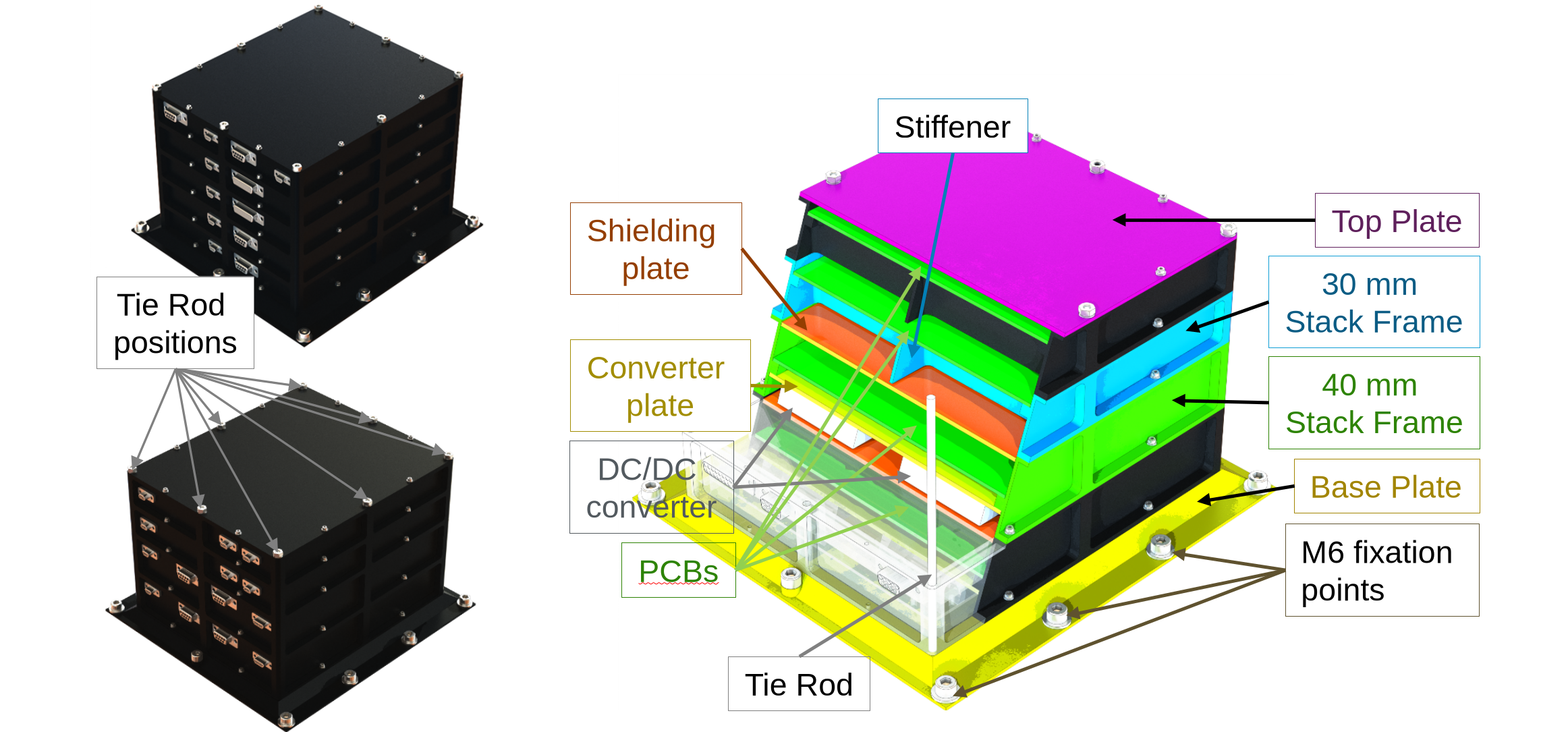}
   \end{center}
   \caption[Crossection of the IRT ICS electronics box] 
   { \label{fig:ICS_box} 
Crossection of the IRT ICS electronics box.}
   \end{figure}

The two ICS units (shown in Figure \ref{fig:ICS_box}) consist of several modularly stacked aluminium frames, each containing a PCB of approximately Double Eurocard size. The PCB frames are held in compression between a cover plate and a mounting plate via eight tie-rods, which run through the edges of the frames. As there will be no active thermal control of the ICS units, the stacking order of the PCB frames has been chosen to optimise the conductive heat dissipation to the SVM platform, with active PCBs alternately separated from each other by their cold-redundant backups. The electrical interfaces to each board are routed via external connectors and harnesses.

\subsection{The Data Handling Unit}
The DHU is the IRT primary power, command, and telemetry interface to the spacecraft. It is responsible for the conditioning and distribution of secondary power for the instrument systems, the control and management of the instrument through tele-commands, automated scheduling and FDIR actions, as well as the processing and on-board analysis of scientific data from the IRT camera.

\noindent The DHU consists of two types of PCBs, each with a cold-redundant backup:

\begin{itemize}
	\item\textbf{Power Supply Board -} The PSB is a power conditioning and distribution board. It recieves the unregulated 28 V primary power from the SVM power supply and generates the lower voltage power supplies required for the IRT subsystems. The PSB implements a combination of SMRT and SVHF DCDC converters in order to provide high conversion efficiency while minimising the volume requirements of the PCB. The PSB is responsible for executing the switch-on sequence of the instrument, and implementing protections against under voltage and over voltage.

	\item\textbf{Data Processing Board -} The DPB performs the core control and data-handling functions of the DHU. It hosts the central processor, local mass memory, time distribution circuits, and housekeeping collection circuits. The central element of the DPB is the Frontgrade Gaisler GR740 microprocessor. The GR740 is a quad-core LEON4FT SPARC V8 processor, which provides the necessary computational power to perform the near-real time processing of the IRT data (described in section 5.3). The DPB also hosts 32 GB of NAND FLASH memory for the storage of an extensive on-board catalogue of known NIR and optical sources used for transient verification and astrometry. Communication between the DPB, the spacecraft OBDH, and the payload module's Master DHU system (through which the automatic observation plan is coordinated) is implemented via an SVM mounted SpaceWire routing switch.

\end{itemize}

\subsection{The Electronics Control Unit}
The ECU is situated between the DHU and the IRT IOS, and consists of PCBs dedicated to the readout and control of the detector, filter wheel, calibration unit, and M2 mirror refocusing subsystems. With the exception of the Detector Control Unit, each PCB has a cold-redundant backup in the ECU. The ECU boards are commanded and monitored by the DHU-DPB using a combination of SpaceWire and Controller Area Network (CAN) interfaces (depending on the requirements of the specific interface).

\begin{itemize}
    \item\textbf{Detector Control Unit -} The DCU implements the functions for the operation, readout, and pre-processing of the IRT detector array. The DCU generates local power supplies/biases for the detectors, while also providing configuration and clocking signals to the SIDECAR ASIC. In observation modes the DCU performs mixed accumulation (MACC) readout (TBC) of the detectors via the FEE ASIC, and performs low level processing on the detector data before formatting and sending the data to the DHU-DPB over SpaceWire. 
    \item\textbf{M2 Thermal Control Board -} In order to compensate for possible misalignment of the telescope mirrors after launch, the mounting truss of the M2 mirror has a series of heaters which allow for fine adjustment of the M1-M2 distance through precise control of the thermal expansion of the truss legs. The ECU's MTCB controls the M2 refocusing mechanism via a closed loop of truss heaters and temperature sensors. 
    \item\textbf{Filter Wheel and Calibration Unit Control Board -} The FWCUCB combines the control and monitoring of both the Filter Wheel and Calibration Unit subsystems in a single PCB. Control of the Filter Wheel Assembly is performed by first identifying the current wheel position by processing the output of the Filter Wheel Assembly's Contactless Position Sensor system, then referencing a local look-up-table to determine the control signal required to actuate the stepper motor to the target position.

    While the IRT is in Calibration mode the FWCUCB activates and controls the Calibration Unit. The FWCUCB uses pulse width modulation to provide one LED at a time with a tuneable current, providing a large dynamic range of calibration source fluxes.
\end{itemize}

\subsection{Onboard Science Software}\label{sec:ics:analysis}
\begin{figure}
  \begin{center}
    \includegraphics[width=14cm]{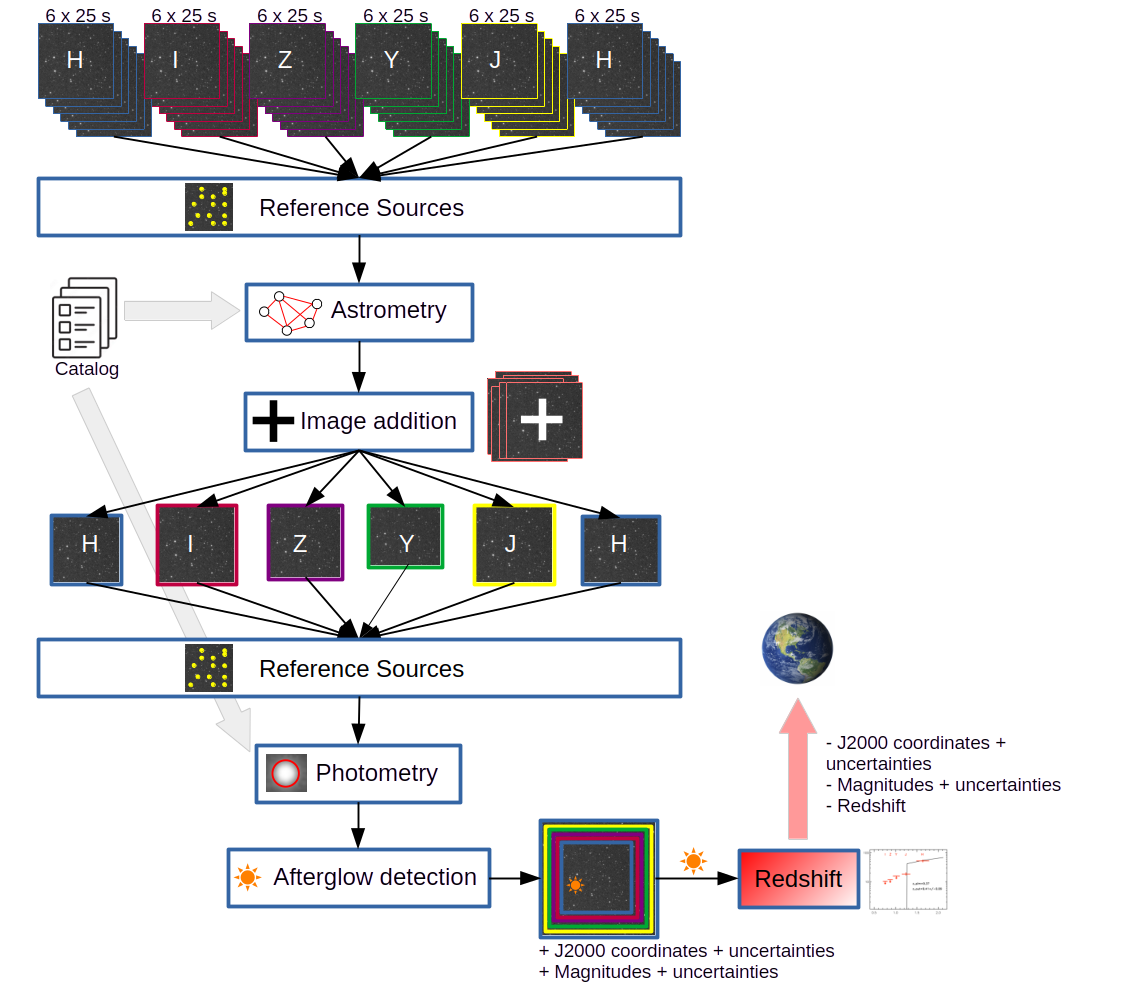}
  \end{center}
  \caption{\label{fig:analysis_sequence}On-board analysis pipeline: see section~\ref{sec:ics:analysis}.}
\end{figure}

One science objective for the IRT is to localize the gamma-ray burst afterglow and to estimate its redshift. This analysis must be conducted on-board to swiftly send the results to the ground. To this end, an analysis pipeline has been designed to run analysis tasks, the technical and scientific performance of which has been evaluated. The pipeline workflow is illustrated in Figure~\ref{fig:analysis_sequence}.

The first stage of the pipeline aims at combining the 6 sub-images collected with the same filter to obtain one single image of 150 s of exposure. This combination must be performed after precisely aligning the 6 images. To that end, an astrometric analysis has to be run on board. It relies on a catalogue of NIR and optical sources which must be stored on-board. The $M$ brightest NIR sources in the field of view are located. They are used to conduct an astrometry analysis to accurately determine the pointing direction associated to the image. For the IRT on-board analysis, the SCAMP method~\cite{2006ASPC..351..112B} was selected as a good compromise to minimize computing resources and maximizing the scientific performance. Indeed, considering the Gaia EDR3 catalog~\cite{2006ASPC..351..112B} and setting $M=20$, we have shown that a pointing accuracy below 2~arcseconds can be achieved for 98~\% of sources. Moreover, the computing time is kept below 8~s which is compliant with the on-board computer characteristics. After the pointing direction is estimated for each image, the 6 images recorded with the same filter can be integrated into one single image, pixel by pixel.

The six images (H, I, Z, Y, J, and H again) are then calibrated in amplitude. A photometric analysis is conducted considering again the $M$ brightest sources in the field of view. The sources previously identified in the catalogue are used to determine the calibration factors and an overall photometric factor is derived from the median of the $M$ sources. In this process, the background component is estimated from an empty region in the image and is subtracted from the image. For IRT, it is required that the photometric accuracy is better than 5~\%. Further studies will be needed to demonstrate this performance number. However, this requirement is typical for a standard photometric analysis.

The gamma-ray burst afterglow is identified as a source not overlapping the known NIR sources in the catalog. It is localized first in the camera plane and then in the sky using the telescope pointing direction determined by the astrometry analysis. The afterglow localization accuracy is impacted by two effects: the accuracy of the point source localization in the camera plane and the stability of the pointing. The former is given by the width of the point spread function and by the stability of the satellite during the 25 s of acquisition. The point spread function of the IRT is expected to concentrate 60~\% of the signal in a circle of 0.9 arcsec radius. The typical pointing stability is better than 1 arcsec. In the end, the overall localization accuracy in the camera plane is 1.4 arcsec. This number is compliant with the IRT requirement.

The redshift of the gamma-ray burst afterglow is evaluated on board by comparing the magnitude measured in the 6 filters and the expected magnitude derived from a synthetic afterglow spectrum. The intrinsic photon flux of the afterglow is modeled as a power-law of time and energy. The extinction from galaxies (resp inter-galactic medium) is modeled in Pei et al.~\cite{1992ApJ...395..130P} (resp.~\cite{2021RNAAS...5..126M}) and has been implemented in Robinet et al.~\cite{snl-repo}. The idea of the photometric redshift analysis is to scan the parameter space to calculate a bank of magnitude templates and to find the best template matching the measured magnitude of the afterglow given its uncertainty. A $\chi^2$ minimization method is used to best fit the data from the six filters. Moreover, we assume a limiting magnitude of 21 beyond which the afterglow cannot be detected. The redshift estimate is mainly obtained by our ability to locate the wavelength position of the Lyman-$\alpha$ break in the spectrum. For this study, we constructed a bank of 62500 templates, where the redshift parameter is logarithmically sampled with 50 values between $z=5.5$ and $z=12$.

\begin{figure}
  \begin{center}
    \includegraphics[width=14cm]{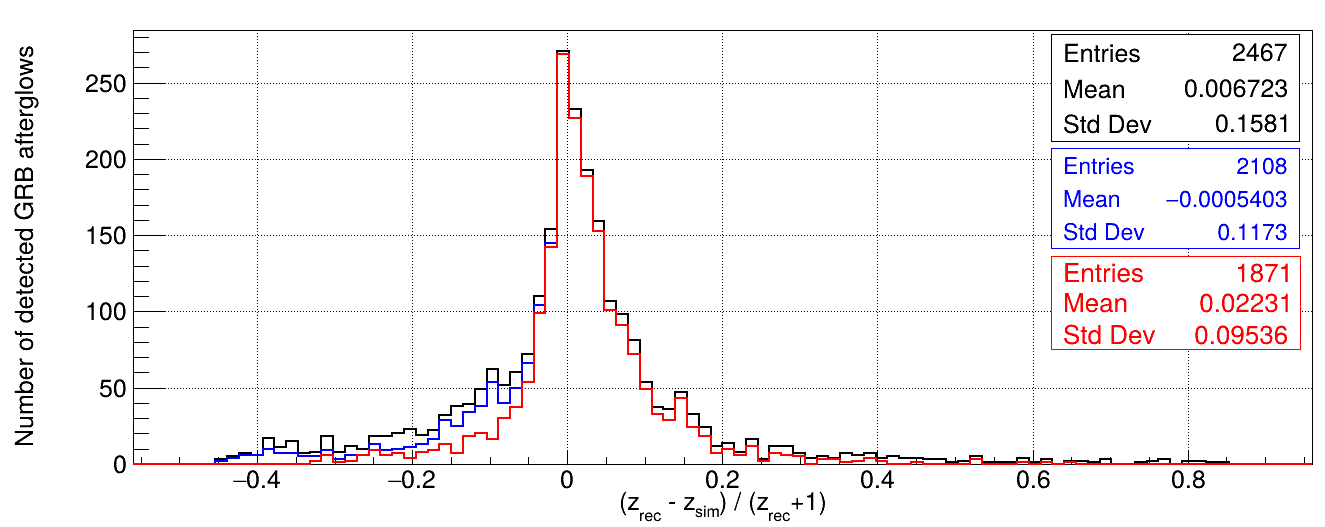}
  \end{center}
  \caption{\label{fig:zres}Resolution for redshift reconstruction for simulated GRB afterglows. The resolution is plotted for afterglows detected with at least one filter (black) and with at least 2 filters (blue). The red histogram shows the redshift resolution for a detection with at least 2 filters and for redshifts larger than 6.}
\end{figure}

To estimate the scientific performance on the redshift reconstruction, we generate time-dependent spectra of gamma-ray burst afterglows using the model described previously (intrinsic flux + extinction). We measure the signal magnitude for each different colour filter and we apply a random jitter of 5~\% corresponding to the required uncertainty for the IRT. We reconstruct the redshift with the $\chi^2$ minimization method and the template bank described above. The resulting redshift estimation is presented in Figure~\ref{fig:zres}. The redshift resolution increases when the signal is detected in multiple colour filters. Indeed, the Lyman-$\alpha$ break is best located when the afterglow is detected in multiple energy bands. When restricting the afterglow sample to $z > 6$, the resolution on the redshift reconstruction is below 10~\% as required by the THESEUS mission design. This result can be further improved by optimizing the template bank. The discretization of the parameter space can be studied to better place the templates to optimize the redshift reconstruction and to reduce the number of templates.

\section{The IRT Performance}

In this section, we present a synthesis of the science performance of the IRT, based on end to end modelling of its photometric and spectroscopic modes that include realistic detector properties, optical performance, line-of-sight (LoS) stability, and background environment. We show that photometric sensitivity requirements are met in all filters (I, Z, Y, J, H) with comfortable SNR margins over the entire allowed jitter–smear stability domain, and that spectroscopic resolving power and SNR requirements are fulfilled throughout the corresponding domain under realistic assumptions on background and readout mode. These performances ensure the IRT capability to provide rapid, accurate GRB localizations, photometry, and redshift estimates up to at least $z \sim$ 10.

The analysis is based on the M7 configuration and end of life (EoL) assumptions, including conservative 20\% margin on top level science performance (SNR and resolving power). The objective is to demonstrate that the instrument robustly satisfies the THESEUS mission requirements.

\subsection{Science Drivers and Main Assumptions}
\subsubsection{Optical and detector configuration}
The baseline IRT configuration includes:
\begin{itemize}
    \item Primary mirror diameter: 0.7 m,
    \item Secondary mirror diameter: 0.21 m,
    \item maging FoV: 15 × 15 arcmin$^{2}$ (vignetting $\leq$ 10\%),
    \item Spectroscopic FoV: 2 × 2 arcmin$^{2}$,
    \item Detector: HAWAII 2RG, 2048 × 2048 pixels, 18 $\mu$m Pixel scale: 0.6 arcsec/pixel,
    \item Photometric filters: I, Z, Y, J, H (standard ground based definitions at this stage),
    \item Spectroscopic coverage: 0.8–1.6 $\mu$m with a grism at the telescope exit pupil, resolving power R $\geq$ 400 at 1.1 $\mu$m.
\end{itemize}

The effective collecting area baseline is approximately 0.35 m$^{2}$, in line with THESEUS telescope requirements. The overall throughput in spectroscopy (telescope + instrument + grism + detector quantum efficiency(QE)) at 1.1 $\mu$m is of order 0.55 at End of Life (EoL) conditions.

\subsubsection{Noise Sources and Background Environment}
\label{sec:backgrounds}
The instrument performance is driven by different noise and background sources that we considered in our model. First of all there is the statistical photon noise from the astronomical source. In addition the other sources of background include the irreducible zodiacal light, in- and out-field stray-light from stars and the Earth limb, and the thermal emission from the telescope and instrument itself. To these we have to add the detector noise, including the readout noise (RoN), assumed $\simeq$ 13 e$^{-}$ rms in Correlated Double Sampling (CDS), and the dark current, assumed $\simeq$ 1 e$^{-}$/s/pixel at $\sim$120 K.

The zodiacal background is modelled using Aldering (2001)\cite{Aldering2001} in photometry and the Leinert map (with the normalization based on the Solar Mass Ejection Imager (SMEI) data \cite{Leinert1998}) in spectroscopy. For baseline performance assessment, an ecliptic latitude of $|\beta|$ = 30$^{\circ}$ is adopted, with the constraint that the Earth limb remains at least 26$^{\circ}$ away from the LoS depending on mode.

Thermal background is evaluated using tools developed for Euclid/NISP, assuming mirror emissivity $\simeq$ 0.03 and instrument emissivity $\simeq$ 0.75. Over 0.7–1.8 $\mu$m, the telescope thermal contribution is negligible compared to zodiacal light for the assumed temperature range, but it becomes non negligible in spectroscopic mode when integrated over 0.8–1.6 $\mu$m.

\subsection{Line of Sight Stability}
The Line of sight (LoS) stability is a major performance driver. In photometry, jitter and drift broaden and smear the PSF, diluting flux over more pixels and decreasing the fraction of flux that falls inside a fixed photometric aperture. This directly lowers SNR at fixed exposure time.
In spectroscopy mode, the same effects elongates the spectral PSF along the dispersion axis, increasing the effective FWHM in wavelength and therefore reducing the resolving power R; it also spreads flux over more pixels, decreasing SNR per spectral element.

The LoS stability model includes:
\begin{itemize}
    \item Random jitter: modelled as Gaussian blurring with typical 3$\sigma$ values $\simeq$ 1 arcsec over 25 s (photometry) and $\simeq$ 1.35 arcsec over 60 s (spectroscopy), in line with THESEUS pointing requirements.
    \item Deterministic drift (“smear”): modelled as a constant velocity motion over a single exposure, leading to convolution of the PSF with a top hat function. Representative values are $\simeq$ 0.011 arcsec/s, with an envelope up to $\simeq$ 0.06 arcsec/s depending on mode.
\end{itemize}

The combined optical PSF is obtained by convolving the diffraction pattern of the obscured 0.7 m pupil with Gaussian kernels representing the telescope and instrument aberrations and platform jitter, plus the induced drift. The final PSF is then sampled by the detector pixel grid.

\subsection{Photometric Performance}
\subsubsection{Method and Simulator}
The photometric performance is evaluated with the THESEUS Infra Red Telescope Photometric Performance Calculator, a dedicated end to end simulator that takes into account the following parameters: the telescope and instrument throughput (including filters), the photometric aperture optimization and image quality metrics (EED50, EED80), the detector properties (pixel size, QE, RoN, dark current), the different background sources (see \ref{sec:backgrounds}), and the line of sight stability (jitter and drift/smear), consistently with the THESEUS M7 pointing requirements.

Further simulation assumptions include the exposure scheme, i.e. 150 s total per filter, split into six 25 s individual frames to limit PSF blurring, on board stacking (images are co added assuming perfect registration; this is conservative since in-flight the photometric aperture will be re optimized based on the measured footprint over 150 s), photometric aperture of 9 pixels (3$\times$3) for point sources, corresponding to $\sim$60–70\% encircled energy depending on band and stability, and CDS readout mode (Fowler and up the ramp (FUR) modes can also be evaluated as options).

The photometric requirements are to reach SNR = 5 in 150 s at 20.9 (AB) for the I band, 20.7 for Z, 20.4 for Y 20.7 for J and 20.8 for H. In the design analysis, a 20\% margin is applied: target SNR = 6 at the requirement magnitude.

\subsubsection{Sensitivity and Dependence on LoS Stability}

Under a representative “standard” LoS stability case (jitter = 0.33 arcsec 1$\sigma$ over 25 s; drift = 0.011 arcsec/s) and background = 1.75 × zodiacal light at $|\beta|$ = 30$^{\circ}$, the simulator shows that all the photometric performances are met with $\geq$ 20\% margin. Even in a degraded stability case (e.g. jitter = 0.45 arcsec 1$\sigma$, drift = 0.03 arcsec/s), the SNR remains $\geq$ 6 in all filters.

\subsubsection{Impact of the readout mode}
In photometry, the noise budget is strongly dominated by Readout Noise (RoN) because individual exposures are only 25 s long. Implementing Fowler or up the ramp sampling (FUR) significantly reduces the effective readout contribution. The FUR with $\sim$5 groups of 3 reads within 25 s yields an SNR gain of about 20\% relative to CDS, while the Fowler sampling (np $>$ 1) offers similar or slightly higher improvements, depending on implementation.

Since the baseline CDS performance already satisfies both the requirement and the goal everywhere in the pointing stability domain, advanced sampling modes are not mandatory for compliance, but provide useful additional margin and may allow, for example, smaller apertures or shorter total exposure time.

\subsection{Spectroscopic Performance}
\subsubsection{Model and Assumptions}
The slit-less spectroscopy performance is evaluated with a dedicated simulator that combines the optical PSF model (diffraction with central obscuration; aberrations modelled as Gaussian convolution),
the platform jitter and drift (smear), with drift direction conservatively aligned along the dispersion axis for the sizing case, a linear dispersion model, tuned such that three pixels sample the FWHM at 1.1 µm (yielding nominal R $\simeq$ 500 at that wavelength), a zodiacal background integrated over 0.8–1.6 $\mu$m, and stray-light and thermal contributions expressed as a factor k of the zodiacal level at $|\beta|$ = 30$^{\circ}$. The detector properties identical to those used for photometry.

The main spectroscopy requirements are R $\geq$ 400 at 1.1 $\mu$m,  SNR = 10 per spectral element at 1.1 $\mu$m in 1800 s for H = 16.4 AB, and SNR = 3 per spectral element at 1.1 $\mu$m in 1800 s for H = 17.5 AB.

A 20\% margin is applied: target R $\simeq$ 480 and SNR $\simeq$ 12 (for the SNR=10 requirement) at EoL. The baseline observing scheme is composed by	30 exposures of 60 s (total 1800 s), stacking assuming accurate registration using zero th order images, and spectral extraction over three detector rows in the spatial direction.

\subsubsection{Resolving Power and Dependence on LOS Stability}
Under nominal assumptions (jitter $\simeq$ 0.45 arcsec 1$\sigma$, drift = 0.011 arcsec/s along the dispersion axis) R $\simeq$ 515 at 1.1 $\mu$m, R $\simeq$ 450 at 0.8 $\mu$m, and R $\simeq$ 700 at 1.6 $\mu$m.

\begin{figure}
  \begin{center}
    \includegraphics[width=10cm]{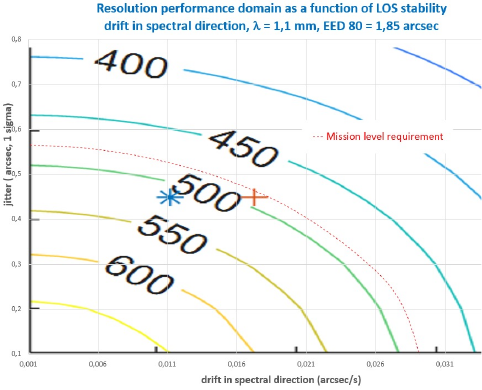}
  \end{center}
  \caption{\label{fig:rvslos}Jitter–smear map of resolving power at 1.1 µm: contour R = 480 (R=400 + 20\% margin) over the THESEUS M7 spectroscopic stability domain.}
\end{figure}

To explore the dependence on LoS stability, a grid of jitter and smear parameters were used to compute R at 1.1 $\mu$m. For each (jitter, drift) pair, a PSF is generated including diffraction, aberrations, jitter and drift over 60 s. The spectral profile of a monochromatic line at 1.1 $\mu$m is extracted and fitted, and the FWHM is measured and $R = \lambda / \Delta\lambda$ is derived. For all combinations of jitter and smear within the mission spectroscopic stability domain, the resolving power at 1.1 $\mu$m. remains $\geq$ 480.
Moving towards larger jitter and larger drift values gradually increases the FWHM and decreases R, but always within the mission requirements. The most constraining direction is when drift is aligned with the dispersion axis; if the drift is mainly perpendicular, the impact on R is significantly smaller.
Thus the resolving power requirement R $\geq$ 400 is met with more than 20\% margin everywhere in the specified jitter–smear domain, see Figure \ref{fig:rvslos}.

\subsubsection{Sensitivity, background and LoS stability}
The spectroscopic SNR at 1.1 $\mu$m depends on the source magnitude (here H = 16.4 or 17.5 AB), the total background at the detector, parametrized as k $\times$ zodiacal background, and the LoS stability (jitter and drift), which influences both the PSF size (signal concentration per pixel) and the extraction aperture needed.

For H = 17.5 AB and H = 16.4 AB, and for a representative point in the jitter–smear domain, at k = 0.7, SNR $\simeq$ 3.6 (H = 17.5) and $\simeq$ 12 (H = 16.4) at 1.1 $\mu$m, after stacking 30 $\times$ 60 s and summing 3 spatial rows. As k increases beyond 0.7, the background noise grows and SNR decreases roughly as 1/$\sqrt{k}$ in the background dominated regime. 
To assess the combined effect of background and LoS stability, SNR was mapped over the jitter–smear plane, for k = 0.7 and k $\simeq$ 0.82, assuming H = 16.4 AB. The study shows that with background controlled at or below 0.7 $\times$ zodiacal, spectroscopic SNR requirements are met with $\geq$ 20\% margin across the full jitter–smear domain, even in CDS. Aslo with slightly higher background (k $\simeq$ 0.82), the required SNR can still be reached across almost the full domain if a more advanced readout mode (FUR) is adopted.

\subsection{Performance Conclusions}

The combined photometric and spectroscopic performance of the THESEUS IRT ensures that the mission’s primary science objectives are met with comfortable margins:
\begin{itemize}
    \item In photometry, SNR requirements are exceeded in all I, Z, Y, J, H bands with about 20\% margin for all jitter–smear conditions allowed at mission level. This supports rapid and accurate GRB afterglow localization, absolute photometry at the few percent level, and robust multi band SED measurements.
    \item In spectroscopy, resolving power R $\geq$ 400 and SNR requirements (10 at H = 16.4, 3 at H = 17.5) are fulfilled over the full jitter–smear domain, provided the background allocation is respected and, in the most demanding configurations. FUR type readout is adopted in this case.
    \item Together, these capabilities enable the IRT to provide fast photometric and spectroscopic redshifts of GRB afterglows up to z$\sim$10, a cornerstone of the THESEUS science case.
\end{itemize}

\section{Conclusions}
We presented the design and performances of the Infra-Red Telescope (IRT) on board the THESEUS mission concept. We have shown that the design studied during the M7 phase A fully responds to the scientific requirements with generally healthy margins, within the resource constraints of the THESEUS mission.
Hence, with such design, IRT will be able to provide in near real-time the identifications and the photometric redshifts for the GRBs detected by THESEUS, and in addition spectroscopically characterise them for the brightest ones.

\subsection{Acknowledgments}
This work is partially supported by the French Space Agency (CNES), by the Italian Space Agency under the Agreement no. 2024-17-HH.0 ASI-INAF, and the Deutsches Zentrum fuer Luft- und Raumfahrt (DLR), grant 50 OO 2404.

\bibliography{report} 
\bibliographystyle{spiebib} 

\end{document}